\documentclass{LMCS}

\def\dOi{9(3:7)2013}
\lmcsheading%
{\dOi}
{1--11}
{}
{}
{Oct.~16, 2012}
{Aug.~29, 2013}
{}

\usepackage{amssymb,enumerate,hyperref}

\newenvironment{definition}{\begin{defi}}{\end{defi}}
\newenvironment{lemma}{\begin{lem}}{\end{lem}}
\newenvironment{theorem}{\begin{thm}}{\end{thm}}

\providecommand{\NN}{\mathbb{N}}
\providecommand{\sm}{\sigma}
\providecommand{\eqdef}{\triangleq}
\providecommand{\seqc}{\mathbf{c}}
\providecommand{\elias}{S}
\providecommand{\continuum}{\mathfrak{c}}
\providecommand{\forcing}{\mathbb{P}}
\providecommand{\tends}{\longrightarrow}
\DeclareMathOperator{\val}{val}
\DeclareMathOperator{\supp}{supp}

\title[Universal codes]{Universal codes of the natural numbers}
\author[Y.~Filmus]{Yuval Filmus}
\address{Department of Computer Science, University of Toronto}
\email{yuvalf@cs.toronto.edu}

\ACMCCS{[{\bf Mathematics of computing}]:  Information theory---Coding theory}

\begin{document}

\begin{abstract}
A code of the natural numbers is a uniquely-decodable binary code of the natural numbers with non-decreasing codeword lengths, which satisfies Kraft's inequality tightly. We define a natural partial order on the set of codes, and show how to construct effectively a code better than a given sequence of codes, in a certain precise sense. As an application, we prove that the existence of a scale of codes (a well-ordered set of codes which contains a code better than any given code) is independent of ZFC.
\end{abstract}

\maketitle

\section{Introduction} \label{sec:introduction}
Lossless coding theory concerns the problem of encoding a sequence of symbols in some alphabet, usually binary. We demand two properties from our codes: they need to be (uniquely) decodable, and they should be concise, that is, the codewords should be short. In this paper we address the following problem: how concise can a coding system for the natural numbers be?

In 1975, Elias~\cite{elias75} considered this problem and constructed a sequence of efficient codes, culminating in the so-called $\omega$-code (almost the same code had been discovered by Levenshtein~\cite{levenshtein68english} in 1968). The third member in Elias's sequence of codes, called the $\gamma$-code, is already asymptotically optimal in the sense that given a non-decreasing high-entropy distribution on the natural numbers, the expected codeword length is almost optimal; consult Elias~\cite{elias75} for a formal definition.

Other concise codes have been constructed by Bentley and Yao~\cite{bentley+yao76}, Even and Rodeh~\cite{even+rodeh78} and Stout~\cite{stout80}. These codes were analyzed by Ahlswede et al.~\cite{ahlswede+97}. More recent examples are Yamamoto~\cite{yamamoto00} and Tarau~\cite{tarau10}. An online universal code has been constructed by Dodis et al.~\cite{dodis+10}.

A natural question to ask is whether there exists an optimal code. We formulate this question in Section~\ref{sec:main} and show that not only is there no single optimal code, but there is also no optimal \emph{sequence} of codes. Since the proofs of these results are constructive, they can be used to construct a fast-growing hierarchy of codes. Elias's construction cannot be used to obtain this result, as we show in Section~\ref{sec:elias}.

Care must be taken when considering the \emph{practical} implications of these results: while all codes we consider are effective, they are not necessarily efficient, in the sense that encoding and decoding could be slow. Furthermore, in practice one is not interested in the asymptotic performance of a code, but in its performance on integers up to a certain application-specific bound, or even on a certain class of distributions.

We go on further and consider the existence of a \emph{scale} of codes, which is an uncountable sequence of codes, ordered so that latter codes are better (in the sense of Definition~\ref{def:code-order} below), and containing a code better than any given code. We show in Section~\ref{sec:scale} that the existence of a scale is independent of ZFC, imitating classical results on functions on the natural numbers ordered by dominance.

\section{Definitions} \label{sec:definitions}
We start with some basic notations. The set of all finite binary strings is denoted $\{0,1\}^*$. The set of natural numbers (including zero) is denoted $\NN$. The set of finite sequences of natural numbers is denoted $\NN^*$. The length of a binary string $x$ is denoted $|x|$.

Next, some terminology from recursion theory.
A sequence $a(n)$ is called \emph{effective} if the mapping $n \mapsto a(n)$ is recursive (computable by an algorithm). A sequence $a_n(m)$ of sequences is effective if the mapping $(n,m) \mapsto a_n(m)$ is recursive. A real number $x$ is effective if there is a recursive function mapping $n$ to a closed rational interval of width at most $1/n$ containing $x$ (all rational intervals appearing in this paper are closed).

A sequence $a(n)$ is effective \emph{relative to} another sequence $b(n)$ if the mapping $n \mapsto a(n)$ is recursive given an oracle for the mapping $n \mapsto b(n)$. The concept of being effective relative to a sequence of sequences or to a real number is defined analogously. Similarly we can extend the definition to cover sequences of sequences and real numbers which are effective relative to other data.

We proceed to define binary codes, which are our main focus of study.
\begin{definition} \label{def:binary-code}
 A (uniquely-decodable) \emph{binary code} of the natural numbers is a mapping $C \colon \NN \rightarrow \{0,1\}^*$ with the property that the function $C^*\colon \NN^* \rightarrow \{0,1\}^*$ defined by $C^*(n_1\ldots n_k) = C(n_1)\ldots C(n_k)$ is injective. If furthermore $|C(n)| \leq |C(m)|$ whenever $n \leq m$, then $C$ is \emph{monotone}.

 A \emph{prefix code} has the additional property that $C(n)$ is not a prefix of $C(m)$ for any $n \neq m$.
\end{definition}

Kraft~\cite{kraft49} and McMillan~\cite{mcmillan56} proved the following well-known inequality.
\begin{lemma}[Kraft's inequality] \label{lem:kraft}
 Let $C$ be a binary code. Then
 \[ \sum_{n \in \NN} 2^{-|C(n)|} \leq 1. \]
 Conversely, given a sequence $c\colon \NN \to \NN$ satisfying the inequality
 \[ \sum_{n \in \NN} 2^{-c(n)} \leq 1, \]
 there exists a prefix code $C$ such that $|C(n)|=c(n)$. Furthermore, $C$ is effective relative to $c$.
\qed
\end{lemma}

Due to this inequality and its converse, our study will concentrate only on the \emph{lengths} of codewords rather than the codewords themselves. This prompts the following definition.

\begin{definition} \label{def:code}
 A \emph{precode} is a monotone non-decreasing function $c \colon \NN \rightarrow \NN$ satisfying Kraft's inequality
 \[ \sm(c) \eqdef \sum_{n=0}^\infty 2^{-c(n)} \leq 1. \]
 A \emph{code} is a precode in which Kraft's inequality is tight. A \emph{proper precode} is a precode in which Kraft's inequality is strict.
\end{definition}
The theory can also be developed with respect to non-monotone codes, but we feel that this is less natural. We require that Kraft's inequality be tight for technical reasons (to make our constructions effective). We feel that this is not a large concession since (as we show in Section~\ref{sec:main}) any binary code can be improved to a binary code in which Kraft's inequality is tight.

Following properties of the sequence of codes constructed by Elias~\cite{elias75}, we define a partial order on precodes.

\begin{definition} \label{def:code-order}
Let $c,d$ be precodes. We say that $c \prec d$ (read \emph{$c$ is better than $d$}) if
\[ \lim_{n \rightarrow \infty} c(n) - d(n) = -\infty. \]
We say that $c \preceq d$ if
\[ \sup_{n \rightarrow \infty} c(n) - d(n) < \infty. \]
\end{definition}

This definition corresponds to the ratio test for convergent series: indeed, with any precode $c$ we can associate a convergent series $c'(n) = 2^{-c(n)}$, and then $c \prec d$ if and only if $c'(n)/d'(n) \rightarrow \infty$. This differs from the definition used by Cholshchevnikova~\cite{cholscevnikova} and Vojt\'a\v{s}~\cite{vojtas}, who apply the ratio test to the remainder term.

Armed with this definition, we can give some evidence to our claim that non-monotone codes are less natural.
\begin{lemma} \label{lem:non-monotone}
There is a function $d\colon \NN \rightarrow \NN$, satisfying Kraft's inequality tightly, such that $c \npreceq d$ for any code $c$.
\end{lemma}
\proof
Define $d$ as follows:
\[
 d(n) = \begin{cases} k + 2 & \text{if } n = 4^k + k - 1, \\ 3k + 2 & \text{if } 4^k + k \leq n \leq 4^{k+1} + k - 1. \end{cases}
\]
The critical values, $n = 4^k + k - 1$, are $0, 4, 17, 66, \ldots$ and so on. Let us check that $d$ satisfies Kraft's equality:
\[
 \sum_{n=0}^\infty 2^{-d(n)} = \sum_{k=0}^\infty \left(2^{-k-2} + 4^k \cdot 2^{-3k-2}\right) = \sum_{k=0}^\infty 2^{-k-1} = 1.
\]
If $c$ is any code then for any $n$ we have
\[ 1 > \sum_{m=0}^n 2^{-c(m)} \geq (n + 1)2^{-c(n)}. \]
Thus $c(n) > \log_2 (n + 1)$. Choosing $n = 4^k + k - 1$, we conclude that
\[ c(4^k + k - 1) > \log_2 (4^k + k) > 2k. \]
Therefore $c(4^k + k - 1) \geq 2k + 1 = d(4^k + k - 1) + k - 1$, and so $\sup_{n \rightarrow \infty} c(n) - d(n) = \infty$, that is $c \npreceq d$.
\qed

\section{Existence of optimal codes} \label{sec:main}

Our goal in this section is to show that there is no optimal code, or even optimal sequence of codes. This is the statement of the following theorem.

\begin{theorem} \label{thm:main}
 For every sequence of codes $(c_n)_{n \in \NN}$ there is a code $d$, effective relative to the sequence, such that $d \prec c_n$ for every $n \in \NN$.
\end{theorem}

Similar results in the related context of fast-growing functions were proved by du~Bois-Reymond~\cite{duboisReymond75} and Hadamard~\cite{hadamard94}. Compared to these results, the main challenges in proving Theorem~\ref{thm:main} are constructing $d$ in an effective way, and ensuring that $d$ is monotone.

The first step in proving Theorem~\ref{thm:main} is constructing effectively a precode $e$ satisfying $e \prec c_n$ for every $n \in \NN$.

\begin{lemma} \label{lem:better}
 For every sequence of codes $\seqc = (c_k)_{k \in \NN}$ there is a proper precode $e$, effective relative to $\seqc$, such that $e \prec c_k$ for every $k \in \NN$. Furthermore, $\sigma(e) \leq 1/2$ and $\sigma(e)$ is also effective relative to $\seqc$.
\end{lemma}
\proof
 Let $d(n) = \min_k c_k(n) + k$. If $k \geq c_0(n)$ then $c_k(n) + k > c_0(n) \geq d(n)$, and so $d(n) = \min_{k < c_0(n)} c_k(n) + k$. This shows that $d$ is effective relative to $\seqc$. Moreover, since the codes $c_k$ are monotone, so is $d$. We will construct a precode $e \prec d$, and it will follow (as we show below) that $e \prec c_k$ for all $k \in \NN$.

 We start by computing a sequence $(p_m)_{m \in \NN}$ satisfying $\sum_{n = p_m}^\infty 2^{-d(n)} \leq 2^{-m}$. For $m,k \in \NN$ let $q_{m,k} \geq 1$ be the minimal index satisfying $\sum_{n = 0}^{q_{m,k} - 1} 2^{-c_k(n)} \geq 1-2^{-m-2}$, and note that $\sum_{n = q_{m,k}}^\infty 2^{-c_k(n)} \leq 2^{-m-2}$. Define $p_m = \max_{k \leq m+1} q_{m,k}$. We have
\begin{multline*}
 \sum_{n = p_m}^\infty 2^{-d(n)} <
 \sum_{n = p_m}^\infty \sum_{k = 0}^\infty 2^{-c_k(n) - k} =
 \sum_{k = 0}^\infty 2^{-k} \sum_{n = p_m}^\infty 2^{-c_k(n)} \\ <
 \sum_{k = m+2}^\infty 2^{-k} + \sum_{k = 0}^{m+1} 2^{-k} \sum_{n = q_{m,k}}^\infty 2^{-c_k(n)} \leq
 2^{-m-1} + \sum_{k=0}^{m+1} 2^{-k} 2^{-m-2} < 2^{-m}.
\end{multline*}
 The existence of $p_0$ implies that $\sigma(d)$ is convergent, and so $d(n) \tends \infty$. Let $J = \{ n \geq 1 : d(n) > d(n-1) \}$. Since $d(n) \tends \infty$, $J$ is infinite.
 The idea now is to construct the sequence $e$ as follows. Choose an appropriate increasing sequence $0 = r_0 < r_1 < \cdots$, and let $e(n) = d(n) - m + C$ for $r_m \leq n < r_{m+1}$. We will choose the points $r_m$ for $m \geq 1$ from the set $J$, and this will ensure that $e$ is monotone. An appropriate choice of the points $r_m$ will ensure that $\sigma(e) < \infty$ is computable (as a function of $C$), and will enable us to choose a value of $C$ guaranteeing $\sigma(e) \leq 1/2$.

 The sequence $(r_m)_{m \in \NN}$ is defined as follows. Let $r_0 = 0$, and for $m \geq 1$, let $r_m$ be the minimal element of $J$ which is larger than both $r_{m-1}$ and $p_{2m}$. The sequence $r$ is clearly effective relative to $\seqc$. Define a sequence $e'$ by $e'(n) = d(n) - m$ in the range $r_m \leq n < r_{m+1}$. The sequence $e'$ is also effective relative to $\seqc$. We claim that $e'$ is monotone. Indeed, if $r_m \leq n < r_{m+1}-1$ then $e'(n+1) = d(n+1) - m \geq d(n) - m = e'(n)$, and if $n = r_{m+1}-1$ then $e'(n+1) = d(n+1) - m - 1 \geq d(n) - m = e'(n)$ since $n + 1 = r_{m+1} \in J$ implies $d(n+1)-1 \geq d(n)$.

 We proceed to show that $\sigma(e')$ is computable. For all $m \in \NN$ we have
\[
 \sum_{n = r_m}^\infty 2^{-e'(n)} = \sum_{l = m}^\infty \sum_{n = r_l}^{r_{l+1}-1} 2^{-d(n) + l} \leq
 \sum_{l = m}^\infty 2^l \sum_{n = p_{2l}}^\infty 2^{-d(n)} \leq
 \sum_{l = m}^\infty 2^l 2^{-2l} = 2^{-m+1}.
\]
 This shows that $\sigma(e')$ is computable. In particular, we can find an integer $C$ such that $\sigma(e') \leq 2^{C-1}$. Define $e(n) = e'(n) + C$. Since $e'$ is monotone so is $e$, and since $\sigma(e) = 2^{-C} \sigma(e') \leq 1/2$, $e$ is a precode. Moreover, $\sigma(e)$ is computable.

 It remains to show that for all $k \in \NN$, $e \prec c_k$. Given $k,t \in \NN$, for all $n \geq r_{k+t+C}$ we have
\[
 e(n) \leq d(n) - k - t \leq c_k(n) - t.
\]
 This implies that $e(n) - c_k(n) \tends -\infty$, and so $e \prec c_k$.
\qed

The second step of the proof of Theorem~\ref{thm:main} completes the precode constructed in Lemma~\ref{lem:better} to a code.
Given a proper precode $e$, we construct a code $d \preceq e$ by pointwise decreasing $e$. The idea is as follows. Suppose that $2^{-k} \leq 1-\sm(e) \leq 2^{-k+2}$. Find the first $m$ such that $e(m) > k$, and create a new code $e'$ by setting $e'(m) = k$ and $e'(n) = e(n)$ for $n \neq m$. The new code satisfies $\sm(e') \geq \sm(e) + 2^{-k-1}$ and so $1-\sm(e') \leq (7/8)(1-\sm(e))$. Repeating this operation, we obtain a code $d$.

The main difficulty is computing an integer $k$ such that $2^{-k} \leq 1-\sm(e) \leq 2^{-k+2}$. This is accomplished by computing an approximation to $\log_2(1-\sm(e))$, a function which is the subject of the following routine technical lemma.

\begin{lemma} \label{lem:log}
 Let $\delta > 0$ be a rational number and let $x$ be a real number satisfying $x \leq 1 - \delta$. Then $\log_2(1-x)$ is effective relative to $x$ and $\delta$.
\end{lemma}
\proof
 Let $\Delta$ be an integer satisfying $\delta \geq 1/\Delta$. The function $f(t) = \log_2(1-t)$ satisfies $-C \Delta \leq f'(t) \leq 0$ for $t \leq 1-1/(2\Delta)$, where $C = 2\log_2 e > 1$. Hence if $I = [a,b] \subseteq [0,1-1/(2\Delta)]$ is an interval of width $\ell$ containing $x$ then $[f(b),f(a)]$ is an interval of width at most $C\Delta \ell$ containing $f(x)$.

 Given non-zero $n \in \NN$, we show how to compute an interval of length at most $1/n$ containing $\log_2(1-x)$, given $\delta$ and an oracle for $x$. We start by computing $\Delta = \lceil 1/\delta \rceil$ and $N = \lceil 2C\Delta n \rceil \geq 2\Delta$. We ask the oracle for a rational interval $[a,b]$ of length at most $1/N$ containing $x$. Since $1/N \leq 1/(2\Delta)$, we have $b \leq x + 1/(2\Delta) \leq (1 - 1/\Delta) + 1/(2\Delta) = 1 - 1/(2\Delta)$. Therefore $[f(b),f(a)]$ is an interval of width at most $C\Delta/N \leq 1/(2n)$ containing $f(x)$. Finally, using a Taylor series expansion we compute rationals $g(a),g(b)$ approximating $f(a),f(b)$ up to $1/(4n)$. The interval $[g(b),g(a)]$ is a rational interval of width at most $1/n$ containing $f(x)$.
\qed

Given this technical lemma, we are able to implement the program described above for the second step of the proof of Theorem~\ref{thm:main}.

\begin{lemma} \label{lem:completion}
 For any proper precode $e$ there is a code $d$, effective relative to $e$ and $\log_2(1-\sm(e))$, such that $d(n) \leq e(n)$ for all $n \in \NN$.
\end{lemma}
\proof
In this proof, whenever we use the term \emph{effective}, we mean effective relative to $e$ and $\log_2(1-\sm(e))$.

We construct a sequence $d_t$ of precodes converging to $d$ (we make this notion precise below). We will ensure that $\sm(d_t) < 1$ and that the sequences $d_t$ and $\log_2(1-\sm(d_t))$ are effective, and furthermore $\sm(d_t)$ is strictly increasing.

The starting point is the sequence $d_0(n) = e(n)$. Next suppose that $d_t$ has been defined. We will find effectively an integer $k_t$ satisfying
\begin{equation} \label{eq:main:2}
 2^{-k_t} \leq 1-\sm(d_t) \leq 2^{-k_t+2}.
\end{equation}
Since $\log_2(1-\sm(d_t))$ is effective, we can effectively find an interval $I_t$ of width at most $1$ containing it, and an integer $k'_t$ such that $I_t \subset [k'_t-2,k'_t]$, implying
\[ 2^{-k'_t} \leq 1-\sm(d_t) \leq 2^{-k'_t+2}. \]
If $t = 0$ then we put $k_t = k'_t$, and otherwise we put $k_t = \max(k'_t,k_{t-1})$. If $k_t = k'_t$ then~\eqref{eq:main:2} clearly holds. If $k_t = k_{t-1}$ then using the assumption $\sm(d_t) > \sm(d_{t-1})$ and the inequality $1-\sm(d_{t-1}) \leq 2^{-k_{t-1}+2}$ we have
\[ 2^{-k_t} \leq 2^{-k'_t} \leq 1 - \sm(d_t) < 1 - \sm(d_{t-1}) \leq 2^{-k_{t-1}+2} = 2^{-k_t+2}. \]

Given $d_t$ and $k_t$, define $d_{t+1}$ as follows. Let $m_t$ be the minimal position for which $d_t(m_t) > k_t$. The new sequence $d_{t+1}$ is obtained from $d_t$ by setting $d_{t+1}(m_t) = k_t$ and $d_{t+1}(n) = d_t(n)$ for $n \neq m_t$; our choice of $m_t$ guarantees that $d_{t+1}$ is monotone. We have
\[ 1 - \sm(d_{t+1}) = 1 - \sm(d_t) - 2^{-k_t} + 2^{-d_t(m_t)}. \]
Since $d_t(m_t) \geq k_t+1$,
\[
 1 - \sm(d_{t+1}) \leq 1 - \sm(d_t) - 2^{-k_t-1} \leq \tfrac{7}{8} (1-\sm(d_t)).
\]
This shows that $1 - \sm(d_t) \tends 0$. Moreover, it implies that $k_t \tends \infty$. Clearly $\sm(d_{t+1})$ is effective. Since
\[
 1 - \sm(d_{t+1}) \geq 2^{-k_t} - 2^{-k_t} + 2^{-d_t(m_t)} = 2^{-d_t(m_t)}, 
\]
applying Lemma~\ref{lem:log}, we see that $\log_2(1-\sm(d_{t+1}))$ is effective.

We define $d(n) = \min_t d_t(n)$. Since $k_t$ is non-decreasing and $k_t \tends \infty$, $d$ is effective. Since each $d_t$ is monotone, so is $d$. Clearly $\sm(d) \geq \sm(d_t)$, hence $1-\sm(d_t) \tends 0$ implies that $\sm(d) \geq 1$. On the other hand, each prefix of $d$ is a prefix of $d_t$ for all sufficiently large $t$. Since each $d_t$ is a precode, we deduce that for all $m \in \NN$, $\sum_{n=0}^m 2^{-d(m)} < 1$, and so $\sm(d) \leq 1$. Put together, $\sm(d)$ is a code.
\qed

We are now ready to prove the main theorem.
\proof[Proof of Theorem \ref{thm:main}]
 Lemma~\ref{lem:better} shows that there is a proper precode $e$ satisfying $e \preceq c_n$ for all $n \in \NN$ which is effective relative to $\seqc$, and furthermore $\sigma(e) \leq 1/2$ is also effective relative to $\seqc$. Lemma~\ref{lem:log} implies that $\log_2(1-\sm(e))$ is effective relative to $\seqc$, and so we can apply Lemma~\ref{lem:completion} to obtain a code $d$ satisfying $d(m) \leq e(m)$ for all $m \in \NN$ which is effective relative to $\seqc$. This clearly implies that $d \prec c_n$ for all $n \in \NN$.
\qed

\subsection{Elias's construction} \label{sec:elias}
The proof of Theorem~\ref{thm:main} is somewhat complicated, and one wonders whether there is any simpler construction. In this section we explain Elias's construction, and show that it doesn't always produce a better code.

Elias~\cite{elias75} defines a sequence of codes, starting with the trivial code $\alpha(n) = n+1$. Successive codes in the sequence are defined by applying the following operation.

\begin{definition}
 Let $c$ be a code. The \emph{successor code} $\elias(c)$ is defined by
 \[ \elias(c)(n) = \lfloor \log_2 (n+1) \rfloor + c(\lfloor \log_2 (n+1) \rfloor). \]
\end{definition}

\begin{lemma} \label{lem:elias}
 For any code $c$, $\elias(c)$ is a code which is effective relative to $c$.
\end{lemma}
\proof
 Clearly $\elias(c)$ is monotone and effective relative to $c$. It also satisfies Kraft's equality:
 \[
  \sum_{n=0}^\infty 2^{-\elias(c)(n)} = \sum_{m=0}^\infty \sum_{n=2^m-1}^{2^{m+1}-2} 2^{-m-c(m)} = \sum_{m=0}^\infty 2^{-c(m)} = 1.
  \eqno{\qEd}
 \]\smallskip

\noindent If we start with $\alpha$ and apply the operation $\elias$ successively, then we obtain progressively better codes. However, this phenomenon isn't universal.

\begin{lemma} \label{lem:elias-bad}
 There exists an effective code $c$ such that $\elias(c) \npreceq c$.
\end{lemma}
\proof
The construction proceeds in infinitely many stages. We start with the empty sequence. Suppose that in stage $n \in \NN$, the sequence is of length $\ell_n$ (so $\ell_0 = 0$). We add to the sequence $2^{\ell_n+1}$ copies of the number $\ell_n + n + 2$. The resulting sequence has the form $2,2,5,5,5,5,5,5,5,5,14,\ldots$ and so on.

The sequence is clearly monotone, and the contribution of stage~$n$ to the sum in Kraft's inequality is $2^{\ell_n+1} \cdot 2^{-\ell_n-n-2} = 2^{-n-1}$. As $\sum_{n \geq 0} 2^{-n-1} = 1$, $c$ is a code. To see that $\elias(c) \npreceq c$, notice that
\[ \elias(c)(2^{\ell_n+1}) = c(\ell_n+1) + \ell_n + 1 = c(\ell_n+2^{\ell_n+1}) + \ell_n + 1 \geq c(2^{\ell_n+1}) + \ell_n + 1. \eqno{\qEd} \]\smallskip

\noindent This lemma shows that Elias's construction cannot be used in place of Lemma~\ref{lem:better}. In the same paper, Elias also defines the $\omega$-code, which is obtained through a diagonalization-like construction from the sequence of codes $\elias^{(t)}(\alpha)$. We do not know how to generalize this construction.

\section{Existence of scale} \label{sec:scale}

In the preceding section, we have shown that there is no optimal sequence of codes. However, if we widen our scope by allowing uncountable sequences, such an object could perhaps be found.

\begin{definition} \label{def:scale}
 A \emph{scale} of codes $S$ is a set which is well-ordered with respect to $\prec$ (every non-empty subset of $S$ has a maximal element) and is cofinal in the poset of codes (for every code $c$ there is a code $d \prec c$ in $S$).
\end{definition}
Instead of insisting that the scale be well-ordered, we could instead ask for it to be a chain (any two elements are comparable). Standard arguments show that if such an object exists then so does a scale.

Mimicking a result of Hausdorff~\cite{hausdorff07a}, we show that a scale exists given that the continuum hypothesis (CH) holds. This follows from Theorem~\ref{thm:main} using a standard argument.

\begin{theorem} \label{thm:scale:ch}
 If CH holds then there exists a scale of codes.
\end{theorem}
\proof
 We construct a scale $S = \{ s_\alpha : \alpha \in \omega_1 \}$ by transfinite induction on $\omega_1$, using the fact that the cardinality of the set of codes is $\continuum = \aleph_1$. Fix an enumeration $(c_\alpha)_{\alpha < \omega_1}$ of all codes. At step $\alpha$, use Theorem~\ref{thm:main} to construct a code $s_\alpha \prec \{ s_\beta : \beta < \alpha \} \cup \{c_\alpha\}$, using the fact that $\alpha$ is countable. By construction, $S$ is well-ordered. Since $s_\alpha \prec c_\alpha$ for any code $c_\alpha$, $S$ is a scale.
\qed

We can also construct a model in which no scale exists. To that end, following a suggestion by Stefan Geschke~\cite{MO32622}, we add $\omega_2$ codes using Cohen's forcing. Theorem~\ref{thm:main} then implies, using standard arguments, that the poset of codes has no scale. Similar arguments appear in Frankiewicz and Zbierski~\cite[II.5]{frankiewicz+zbierski}, Jech~\cite[\S 24]{jech} and Scheepers~\cite{scheepers93}.

The construction uses the concept of \emph{code prefix}, which represents partial information regarding a code.

\begin{definition} \label{def:code-prefix}
 A \emph{code prefix} is a finite non-decreasing sequence $c(0),\ldots,c(n)$ of natural numbers satisfying Kraft's inequality strictly, $\sm(c) < 1$. 
 
 We say that a code (or code prefix) $d$ \emph{extends} a code prefix $c$ if, as a sequence, $c$ is a prefix of $d$.
\end{definition}

The following lemma encapsulates all the information we need to know about codes, gleaned mainly from Theorem~\ref{thm:main}.

\begin{lemma} \label{lem:code-prefixes}
 Let $c$ be a code prefix.
\begin{enumerate}[\em(a)]
 \item The code prefix $c$ can be extended to a code in infinitely many ways.
 \item Given any code $d$ and $n \in \NN$, the code prefix $c$ can be extended to a code prefix $b$ such that $b(m) \leq d(m)$ for some $m \geq n$.
\end{enumerate}
\end{lemma}
\proof
 For the first item, let $c=c(0),\ldots,c(n)$ be a code prefix. We can extend $c$ to a code prefix $c(0),\ldots,c(n+1)$ in infinitely many ways. Any such extension can be extended to a code prefix $c'$ such that $\sm(c') = 1-2^{-c(n+1)}$. Finally, extend $c'$ to a code by affixing $c(n+1)+1,c(n+1)+2,\ldots$ at its end.

 For the second item, let $c=c(0),\ldots,c(r)$ be a code prefix, and let $d$ be a code. Since $c$ is monotone, $\sm(c) = A/2^{c(r)}$ for some integer $A$, and so $\sm(c) \leq 1-2^{-c(r)}$. Use Theorem~\ref{thm:main} (with $c_n = d$ for all $n \in \NN$) to construct a code $e \prec d$. Find a point $m \geq \max(n,r+1)$ such that $e(m) \leq d(m) - c(r)$. Extend $c$ by $e(r+1)+c(r),\ldots,e(m)+c(r)$ to form a new sequence $b$. Since $\sm(c) \leq 1-2^{-c(r)}$ and $2^{-e(r+1)} + \cdots + 2^{-e(m)} < 1$, this results in a code prefix, which satisfies $b(m) = e(m) + c(r) \leq d(m)$.
\qed

We are now in a position to describe the forcing construction. The entire construction takes place inside a countable transitive model $M$ of ZFC.

\begin{definition} \label{def:forcing}
 A \emph{code prefix bundle} is an $\omega_2$-sequence of code prefixes, only finitely many of which have non-zero length. The forcing $\forcing$ consists of the set of code prefix bundles, ordered by $c < d$ whenever for each $\alpha < \omega_2$, $c_\alpha$ extends $d_\alpha$.

 The \emph{support} of a code prefix bundle $c$, denoted $\supp c$, is the set of $\alpha < \omega_2$ such that $c_\alpha$ has non-zero length. The support is always finite.
\end{definition}

\begin{lemma} \label{lem:ccc}
The forcing $\forcing$ satisfies the countable chain condition: every antichain in $\forcing$ (a subset $C \subseteq \forcing$ in which any two $c,d \in C$ are incompatible: there is no $e \in \forcing$ satisfying $e < c$ and $e < d$) is at most countable.
\end{lemma}
\proof
Suppose that $C$ is an uncountable antichain in $\forcing$. Since the support of any code prefix bundle is finite, the $\Delta$-system lemma shows that there is an uncountable subset $D \subseteq C$ and a finite subset $S \subseteq \omega_2$ such that $\supp c \cap \supp d = S$ for all $c,d \in D$. For each $\alpha \in S$ there are only countably many possible code prefixes, and so since $S$ is finite, there is an uncountable subset $E \subseteq D$ such that $c_\alpha = d_\alpha$ for all $\alpha \in S$ and $c,d \in E$. However, since $\supp c \cap \supp d = S$ and $c,d$ agree on $S$ for all $c,d \in E$, all code prefix bundles in $E$ are compatible, contradicting the assumption that $C$ is an antichain. 
\qed

Let $G$ be a generic filter over $\forcing$, and construct the model $M[G]$, which contains $G$. Since $\forcing$ satisfies the countable chain condition, the forcing preserves cardinals. In the remainder of the section, we show that $M[G]$ contains no scale of codes.

We start with some consequences of Lemma~\ref{lem:code-prefixes}.

\begin{lemma} \label{lem:G}
 Let $\seqc$ be the $\omega_2$-sequence defined by $c_\alpha = \bigcup_{f \in G} f_\alpha$.
\begin{enumerate}[\em(a)]
 \item For each $\alpha < \omega_2$, $c_\alpha$ is a code. Moreover, for $\alpha \neq \beta$, $c_\alpha \neq c_\beta$.
 \item Every code in $M[G]$ has a name in $M^{\forcing}$ which depends on countably many coordinates of $\seqc$.
 \item Let $d \in M^{\forcing}$ be a name of a code which does not depend on $c_\alpha$. Then $\val(d,G) \nprec c_\alpha$.
\end{enumerate}
\end{lemma}
\proof
 The first item follows directly from Lemma~\ref{lem:code-prefixes}(a).

 The second item follows from the countable chain condition. Indeed, every code $c \in M[G]$ (represented as a set of pairs $(n,c(n))$) has a nice name of the form $\{((n,m),a) : a \in A_{n,m}\}$, where each $A_{n,m} \subseteq \forcing$ is an antichain. Lemma~\ref{lem:ccc} shows that each $A_{n,m}$ is countable, and so $C = \bigcup_{n,m \in \NN} A_{n,m}$ is countable. Each $a \in C$ has finite support, and so altogether the name depends on countably many coordinates of $\seqc$.

 To prove the third item, we show that given $n \in \NN$, any code prefix bundle $f$ can be extended to a code prefix bundle $g$ that forces $c_\alpha(m) \leq d(m)$ for some $m \geq n$. Let $D = \val(d,G)$. Using Lemma~\ref{lem:code-prefixes}(b), we can extend $f_\alpha$ to $h_\alpha$ which satisfies $h_\alpha(m) \leq D(m)$ for some $m \geq n$. The value of the prefix $D(0),\ldots,D(m)$ is forced by some code prefix bundle $k$ extending $f$. Since $d$ doesn't depend on the coordinate $\alpha$, we can assume that $k_\alpha = f_\alpha$. The code prefix bundle $g$ extends $k$ by $g_\alpha = h_\alpha$, and by construction it forces $c_\alpha(m) \leq d(m)$.
\qed

Lemma~\ref{lem:G} allows us to show that the bounding number of the poset of codes is $\omega_1$ while its dominating number is $\omega_2$, implying that there is no scale of codes.

\begin{theorem} \label{thm:no-scale}
 In $M[G]$ there is no scale of codes.
\end{theorem}
\proof
 Let $\seqc$ be the $\omega_2$-sequence defined by $c_\alpha = \bigcup_{f \in G} f_\alpha$.
 Suppose $S$ is a scale of codes. For $\alpha < \omega_1$, let $s_\alpha \in S$ satisfy $s_\alpha \prec c_\alpha$. We claim that $S' = \{ s_\alpha : \alpha < \omega_1 \}$ is cofinal in the poset of codes. Otherwise, there exists a code $s \in S$ such that $s \prec s_\alpha \prec c_\alpha$ for all $\alpha < \omega_1$. Yet according to Lemma~\ref{lem:G}(b), such a code has a name which depends only on countably many coordinates of $\seqc$. Considering any other coordinate $\alpha < \omega_1$, Lemma~\ref{lem:G}(c) shows that $s \nprec c_\alpha$.

The fact that $S'$ is cofinal contradicts Lemma~\ref{lem:G}(c) in a different way: according to Lemma~\ref{lem:G}(b), all codes in $S'$ have names depending (together) on at most $\omega_1$ coordinates of $\seqc$. Considering any other coordinate $\alpha < \omega_2$, Lemma~\ref{lem:G}(c) shows that $s \nprec c_\alpha$ for all $s \in S'$, contradicting the fact that $S'$ is cofinal. We conclude that $S$ cannot have been a scale.
\qed

\section{Discussion} \label{sec:discussion}
\textbf{Fast-growing hierarchies.} Theorem~\ref{thm:main} can be used to construct a fast-growing hierarchy of effective codes. Let $\mu$ be a countable ordinal, and assign a computable fundamental sequence $(\alpha^{(i)})_{i \in \NN}$ to every limit ordinal $\alpha < \mu$. The fast-growing hierarchy $(c_\alpha)_{\alpha<\mu}$ is defined according to the following rules. The base case is $c_0(n) = n+1$. For a successor ordinal $\alpha+1$, use Theorem~\ref{thm:main} to construct a code $c_{\alpha+1} \prec c_\alpha$. For a limit ordinal $\alpha$, use Theorem~\ref{thm:main} to construct a code $c_\alpha$ such that $c_\alpha \prec c_{\alpha^{(i)}}$ for all $i \in \NN$.

\textbf{Cardinal characteristics of the continuum.} Section \ref{sec:scale} shows that the existence of a scale of codes is independent of ZFC. However, a more satisfying answer will explain how this phenomenon is related to other cardinal characteristics of the continuum. Specifically, it is known that if we do not require our codes to be monotone, then the resulting poset of codes is Tukey-equivalent to the ideal of measure-zero sets~\cite[Lemma 4.12]{survey}. Todor\v{c}evi\'{c}~\cite{todorcevic} conjectures that our poset is also Tukey-equivalent to the same ideal.

\end{document}